\documentstyle[aps,eqsecnum,preprint,floats,epsf,epsfig]{revtex}
\begin{document}
\def\be{\begin{eqnarray}}
\def\en{\end{eqnarray}}
\def\non{\nonumber}
\def\la{\langle}
\def\ra{\rangle}
\def\nc{N_c^{\rm eff}}
\def\vp{\varepsilon}
\def\drho{\bar\rho}
\def\deta{\bar\eta}
\def\etap{{\eta^{(')}}}
\def\vma{{_{V-A}}}
\def\vpa{{_{V+A}}}
\def\smp{{_{S-P}}}
\def\spp{{_{S+P}}}
\def\J{{J/\psi}}
\def\ov{\overline}
\def\Lqcd{{\Lambda_{\rm QCD}}}
\def\pr{{\sl Phys. Rev.}~}
\def\prl{{\sl Phys. Rev. Lett.}~}
\def\pl{{\sl Phys. Lett.}~}
\def\np{{\sl Nucl. Phys.}~}
\def\zp{{\sl Z. Phys.}~}
\def\lsim{ {\ \lower-1.2pt\vbox{\hbox{\rlap{$<$}\lower5pt\vbox{\hbox{$\sim$}
}}}\ } }
\def\gsim{ {\ \lower-1.2pt\vbox{\hbox{\rlap{$>$}\lower5pt\vbox{\hbox{$\sim$}
}}}\ } }

\font\el=cmbx10 scaled \magstep2{\obeylines \hfill BNL-HET-01/16}

\vskip 1.5 cm

\centerline{\large\bf Semi-inclusive $B$ Decays and Direct CP
Violation} \centerline{\large\bf in QCD Factorization}
\bigskip
\centerline{\bf Hai-Yang Cheng$^{1,2}$ and Amarjit Soni$^{2}$}
\medskip
\centerline{$^1$ Institute of Physics, Academia Sinica}
\centerline{Taipei, Taiwan 115, Republic of China}
\medskip
\centerline{$^2$ Physics Department, Brookhaven National
Laboratory} \centerline{Upton, New York 11973}
\medskip

\bigskip
\bigskip
\centerline{\bf Abstract}
\bigskip
{\small We have systematically investigated the semi-inclusive $B$
decays $B\to MX$, which are manifestations of the quark decay $b
\to Mq$, within the framework of QCD-improved factorization. These
decays are theoretically clean and have distinctive experimental
signatures. We focus on a class of these that do not require any
form factor information and therefore may be especially suitable
for extracting information on the angles $\alpha$ and $\gamma$ of
the unitarity triangle. The nonfactorizable effects, such as
vertex-type and penguin-type corrections to the two-body $b$
decay, $b\to Mq$, and hard spectator corrections to the 3-body
decay $B\to Mq_1\bar q_2$ are calculable in the heavy quark limit.
QCD factorization is applicable when the emitted meson is a light
meson or a charmonium.  We discuss the issue of the CPT constraint
on partial rate asymmetries. The strong phase coming from
final-state rescattering due to hard gluon exchange between the
final states can induce large rate asymmetries for tree-dominated
color-suppressed modes $(\pi^0,\rho^0,\omega)X_{\bar s}$. The
nonfactorizable hard spectator interactions in the 3-body decay
$B\to Mq_1\bar q_2$, though phase-space suppressed, are extremely
important for the tree-dominated modes
$(\pi^0,\rho^0,\omega)X_{\bar s},~\phi X$, $\J X_s,\J X$ and the
penguin-dominated mode $\omega X_{s\bar s}$. In fact, they are
dominated by the hard spectator corrections. This is because the
relevant hard spectator interaction is color allowed, whereas the
two-body semi-inclusive decays for these modes are
color-suppressed.  Our result for ${\cal B}(B\to \J X_s)$ is in
agreement with experiment. The semi-inclusive decay modes: $\ov
B^0_s\to (\pi^0,\rho^0,\omega)X_{\bar s}$, $\rho^0X_{s\bar s}$,
$\ov B^0\to(K^-X,K^{*-}X)$ and $B^-\to (K^0X_s,K^{*0}X_s)$ are the
most promising ones in searching for direct CP violation. In fact,
they have branching ratios of order $10^{-6}-10^{-4}$ and CP rate
asymmetries of order $(10-40)\%$. The decays $\ov
B^0_s\to(\pi^0,\rho^0,\omega)X_{s\bar s}$ and $B^-\to\phi X$ are
electroweak-penguin dominated. Some of them have sizable branching
ratios and observable CP asymmetries.

}

\pagebreak

\section{Introduction}

Recently Beneke, Buchalla, Neubert and Sachrajda
(BBNS)\cite{BBNS1} have proposed a QCD-improved factorization
approach to a class of exclusive $B$-decays that appears quite
promising. BBNS suggest that nonfactorizable effects in an
exclusive decay $B\to M_1M_2$ with recoiled $M_1$ and emitted
light meson $M_2$ are calculable since only hard interactions
between the $(BM_1)$ system and $M_2$ survive in the heavy quark
limit. In this paper we extend the application of BBNS idea of QCD
factorization to a certain class of semi-inclusive decays. In this
regard our approach complements the recent works of He {\it et al}
\cite{He,Hephi}. The semi-inclusive decays that are of special
interest originate from the quark level decay, $b \rightarrow M +
q$; they are theoretically cleaner compared to exclusive decays
and have distinctive experimental signatures \cite{Soni}. Since
these semi-inclusive decays also tend to have appreciably larger
branching ratios compared to their exclusive counterparts, they
may therefore be better suited for extracting CKM-angles and for
testing the Standard Model (SM).

Earlier studies of semi-inclusive decays are based on naive
factorization \cite{Soni} or generalized factorization
\cite{Browder} in which nonfactorizable effects are treated in a
phenomenological way by assuming that, for example, the number of
colors is a free parameter to be fitted to the data. Apart from
the unknown nonfactorizable corrections, this approach encounters
another major theoretical uncertainty, namely the gluon's
virtuality $k^2$ in the penguin diagram is basically unknown,
rendering the predictions of CP asymmetries not trustworthy.

The aforementioned difficulties with the conventional methods can
be circumvented in the BBNS approach of QCD-improved factorization
(QCDF). Indeed, by placing an energy cut $E_M \geq {\rm 2.1~GeV}$
on the meson in the semi-inclusive signal $B \to M + X$, not only
we can enhance the presence of the two body quark decays, $b \to M
+ q$, but also $M$ then can play the role of $M_2$ and $X$ that of
the recoiled meson $M_1$ in the above-mentioned exclusive decay $B
\to M_1 + M_2$, in so far as considerations of BBNS go.
Furthermore, a very important theoretical simplification occurs in
the semi-inclusive decays over the exclusive decays if we focus on
final states such that $M$ does not contain the spectator quark of
the decaying $B(B_s)$ meson as then we completely by-pass the need
for the transition form factor for $B(B_s) \to M$. Recall that for
the exclusive case, in general, we need a knowledge of two such
form factors if $M$ is a pseudoscalar meson or of four form
factors if $M$ is a vector meson.

The consideration of these semi-inclusive $B$ decays has several other
theoretical advantages over the exclusive ones as well.
For one thing, there is no troublesome infrared
divergent problem occurred at endpoints when working in QCD
factorization. As for $CP$ violation, contrary to the exclusive
hadronic decays, it is not plagued by the unknown soft phases.
Consequently, the
predictions of the branching ratios and partial rate asymmetries
for $B\to MX$ are considerably clean and reliable.

Recently QCD factorization has been applied to charmless
semi-inclusive decays $B\to K(K^*)X$ and $B\to\phi X_s$ in
\cite{He,Hephi}. In the present paper we will systematically study
all semi-inclusive decays $B\to PX(X_s,X_c)$, $B\to VX(X_s,X_c)$ and
$B\to\J X$ with $P~(V)$ being a light pseudoscalar (vector) meson,
$X$ the final state containing no charmed or strange particles and
$X_s$ the final state containing a strange quark
and $X_c$ the final state containing a charm quark. We differ from
\cite{He,Hephi} in two main aspects: First, we have included the
complete twist-3 corrections to the penguin coefficient $a_6$ to
be introduced below and electroweak penguin-type corrections to
the coefficients $a_{7-10}$ induced by tree 4-quark operators;
both have been neglected in \cite{He,Hephi}. Second, the hard
spectator interaction in the 3-body decay $B\to Mq_1\bar q_2$ is
neglected in \cite{He,Hephi}, whereas we will show that it is
quite important for color-suppressed modes. As a by-product, we
shall see that the troublesome infrared divergent problem
encountered in the exclusive two-body decays does not occur in the
semi-inclusive case.

The consideration of semi-inclusive decays $B\to MX$ has two more
complications than the corresponding two-body decay, $b\to Mq$.
First,  in the free quark approximation,
the fact that $B\to MX$ can be viewed
as the free $b$ quark decay $b\to Mq$ is justified only in the
heavy quark limit. For the finite $b$ quark mass, it becomes
necessary to consider the initial $b$ quark bound state effect.
This has been investigated recently in \cite{He} using two
different approaches which we will follow closely. Second, the
3-body decay $\ov B\to M+q_1+\bar q_2$ with the quark content
$(b\bar q_2)$ for the $\ov B$ meson could be important for
color-suppressed modes as just mentioned before. Here one needs a
hard gluon exchange between the spectator quark $\bar q_2$ and the
meson $M$ in order to ensure that the outgoing $\bar q_2$ is hard.

In passing, we briefly recall that for experimental purposes a
useful feature of these semi-inclusive decays, $B \to M + X$, with
an energetic $M$, is that these events should have relatively low
multiplicity~\cite{Soni}.

The paper is organized as follows. In Sec. II we will outline the
QCD factorization approach relevant for our purposes and then we
proceed to apply it to the two-body decays of the $b$ quark $b\to
Mq$ in Sec. III. In Sec. IV we study the hard spectator
corrections and summarize the results for branching ratios and CP
rate asymmetries. Sec. V gives our conclusions.

\section{QCD factorization}
In this section we want to suggest that the idea of QCD
factorization \cite{BBNS1} can be extended to the case of
semi-inclusive decays, $B \to M + X$, with rather energetic meson
$M$, say $E_M \geq 2.1~{\rm GeV}$. Recall that it has been shown
explicitly \cite{BBNS1} that if the emitted meson $M_2$ is a light
meson or a quarkonium in the two-body exclusive decay $B\to
M_1M_2$ with $M_1$ being a recoiled meson, the transition matrix
element of an operator $O$, namely $\la M_1M_2|O|B\ra$, is
factorizable in the heavy quark limit. Schematically one has
\cite{BBNS1}
 \be
 \la M_1M_2|O_i|B\ra &=& \la M_1M_2|O_i|B\ra_{\rm fact}
 \left[1+\sum r_n\alpha_s^n+{\cal O}({\Lqcd\over
m_b})\right]  \non \\ &=& \sum_jF_j^{BM_1}(m_2^2)\int^1_0 du\,
T_{ij}^I(u)\Phi_{M_2}(u) \non
\\ && +\int^1_0 d\xi \,du\,dv
\,T^{II}_i(\xi,u,v)\Phi_B(\xi)\Phi_{M_1}(u)\Phi_{M_2}(v),
\label{qcdf}
 \en
where $F^{BM_1}$ is a $B-M_1$ transition form factor, $\Phi_M$ is
the light-cone distribution amplitude, and $T^I,~T^{II}$ are
perturbatively calculable hard scattering kernels. In the naive
factorization approach, $T^I$ is independent of $u$ as it is
nothing but the meson decay constant. However, large momentum
transfer to $M_2$ conveyed by hard gluon exchange implies a
nontrivial convolution with the distribution amplitude
$\Phi_{M_2}$. The second hard scattering function $T^{II}$, which
describes hard spectator interactions, survives in the heavy quark
limit when both $M_1$ and $M_2$ are light or when $M_1$ is light
and $M_2$ is a quarkonium \cite{BBNS1}. The factorization formula
(\ref{qcdf}) implies that naive factorization is recovered in the
$m_b\to\infty$ limit and in the absence of QCD corrections.
Nonfactorizable corrections are calculable since only hard
interactions between the $(BM_1)$ system and $M_2$ survive in the
heavy quark limit.

As an illustrative example of a case when QCDF is not applicable,
let us mention the decay $\ov B^0\to \pi^0 D^0$. QCDF does not
work here because the emitted meson $D^0$ is heavy so that it is
neither small (with size of order $1/\Lqcd$) nor fast and cannot
be decoupled from the $(B\pi)$ system. This is also ascribed to
the fact that the soft interaction between $(B\pi)$ and the $c$
quark of the $D^0$ meson is not compensated by that between
$(B\pi)$ and the light spectator quark of the charmed meson.

As for the semi-inclusive decay $B\to MX$, a momentum cutoff
imposed on the emitted light meson $M$, say $p_M>2.1$ GeV, is
necessary in order to reduce contamination from the unwanted
background and ensure the relevance of the two-body quark decay
$b\to Mq$. For example, an excess of $K(K^*)$ production in the
high momentum region, $2.1< p_{K(K^*)}<2.7$ GeV, will ensure that
the decay $B\to K(K^*)X$ is not contaminated by the background
$b\to c$ transitions manifested as $B\to D(D^*)X\to K(K^*)X'$ and
that it is dominated by the quasi two-body decay $b\to K(K^*)q$
induced from the penguin process $b\to sg^*\to sq\bar q$ and the
tree process $b\to u\bar u s$.  As we shall see shortly, the
kinematic consideration here will restrict the possible forms of
factorization for the matrix element $\la MX|O|B\ra$. By the same
physical argument as in the exclusive case and by treating $M_1=X$
and $M_2=M$, it is natural to expect that the factorization
formula (\ref{qcdf}) can be generalized to the semi-inclusive
decay:
 \be
 \la MX|O|B\ra &=& \la MX|O|B\ra_{\rm
fact}\left[1+\sum r_n\alpha_s^n+{\cal O}({\Lqcd\over m_b})\right]
\non \\ &=& \int^1_0 du\, T^I(u)\Phi_{M}(u)  +\int^1_0 d\xi \,du
\,T^{II}(\xi,u)\Phi_B(\xi)\Phi_{M}(u). \label{qcdfsemi}
 \en

In comparing this equation to the exclusive case Eq. (\ref{qcdf}),
a crucial simplification that has occurred is that the
semi-inclusive case does not involve any transition form
factor(s). This attractive feature holds so long as the meson $M$
does not contain the spectator quark in the initial $B$ meson.
Since lack of knowledge of these form factors is often a serious
limitation in quantitative applications, this adds to the appeal
of the semi-inclusive case. Note also that when the emitted meson
$M$ is a light meson or a quarkonium, the nonfactorizable
corrections to naive factorization are infrared safe in the heavy
quark limit and hence calculable. However, by the same token as
the $\ov B^0\to \pi^0 D^0$ decay, the above QCD factorization
formula is also not applicable to $\ov B^0\to D^0(\ov D^0)X$. The
analog of $\ov B^0\to D^+\pi^-$ in semi-inclusive decays is $\ov
B^0\to\pi^- X_c$. However, there is one difference between them,
namely the hard spectator interaction proportional to the kernel
$T^{II}$ vanishes in the decay $\ov B^0\to D^+ \pi^-$, while it
survives in the 3-body decay $\ov B^0\to \pi^- c\,\bar q$, where
$\bar q$ is the spectator quark of the $B$ meson. This is because
the spectator quark in the $B$ and $D$ bound states in $B-D$
transition is very soft in the heavy quark limit, whereas the same
quark $\bar q$ appearing in the 3-body decay has to be hard so
that a hard gluon exchange between $\pi$ and $\bar q$ is needed.

The factorizable hadronic matrix element $\la MX|O|B\ra$ has the
general expression: \be \la MX|O|B\ra_{\rm fact}=\la M|j_1|0\ra\la
X|j_2|B\ra +\la X|j_1'|0\ra\la M|j_2'|B\ra, \label{fm.e.} \en
where $j_1'$ and $j_2'$ arise from the Fierz transformation of the
operator $O$ and the annihilation term $\la MX|j_1|0\ra\la
0|j_2|B\ra$ is suppressed by order $\Lqcd/m_b$ and hence it will
not be included in Eq. (\ref{fm.e.}). As stressed in \cite{He},
Eq. (\ref{fm.e.}) is not the only way for factorization; there are
other ways of factorization, for example, $\la X_1M|j_1|0\ra\la
X_1'|j_2|B\ra$ with $X_1+X_1'=X$. For the aforementioned
quasi-two-body decay, it seems quite plausible to expect that the
configuration $\la X_1M|j_1|0\ra\la X_1'|j_2|B\ra$ is dominated by
$\la M|j_1|0\ra\la X|j_2|B\ra$ \cite{He};
at least in the perturbative region, the
production of additional $X_1$ in the final state is
suppressed by powers of $\alpha_s$
\cite{He}.

The effective Hamiltonian relevant for hadronic semi-inclusive $B$
decays of interest has the form
\be
{\cal H}_{\rm eff}(\Delta B=1) &=& {G_F\over\sqrt{2}}\Bigg\{
V_{ub}V_{uq}^*
\Big[c_1(\mu)O^u_1(\mu)+c_2(\mu)O^u_2(\mu)\Big]+V_{cb}V_{cq}^*
\Big[c_1(\mu)O^c_1(\mu)+c_2(\mu)O^c_2(\mu)\Big] \non
\\ && \qquad
-V_{tb}V_{tq}^*\left(\sum^{10}_{i=3}c_i(\mu)O_i(\mu)+c_g(\mu)
O_g(\mu)\right) \non \\ && \qquad +V_{cb}V_{uq}^*
\Big[c_1(\mu)\tilde O_1(\mu)+c_2(\mu)\tilde O_2(\mu)\Big]
\Bigg\}+{\rm h.c.},
\en
where $q=d,s$ and
\be
&& O^u_1= (\bar ub)_\vma(\bar qu)_\vma, \qquad\qquad\qquad\qquad~~
O^u_2 = (\bar u_\alpha b_\beta)_\vma(\bar q_\beta u_\alpha)_\vma,
\non \\  && O^c_1= (\bar cb)_\vma(\bar qc)_\vma,
\qquad\qquad\qquad\qquad~~~ O^c_2 = (\bar c_\alpha
b_\beta)_\vma(\bar q_\beta c_\alpha)_\vma, \non \\ && \tilde O_1=
(\bar cb)_\vma(\bar qu)_\vma, \qquad\qquad\qquad\qquad~~~ \tilde
O_2 = (\bar c_\alpha b_\beta)_\vma(\bar q_\beta u_\alpha)_\vma,
\non \\ && O_{3(5)}=(\bar qb)_\vma\sum_{q'}(\bar
q'q')_{\vma(\vpa)}, \qquad \qquad O_{4(6)}=(\bar q_\alpha
b_\beta)_\vma\sum_{q'}(\bar q'_\beta q'_\alpha)_{ \vma(\vpa)},
\\ && O_{7(9)}={3\over 2}(\bar qb)_\vma\sum_{q'}e_{q'}(\bar
q'q')_{\vpa(\vma)}, \qquad O_{8(10)}={3\over 2}(\bar q_\alpha
b_\beta)_\vma\sum_{q'}e_{q'}(\bar q'_\beta
q'_\alpha)_{\vpa(\vma)}, \non  \\ && O_g=\,{g_s\over
8\pi^2}\,m_b\bar q\sigma^{\mu\nu}G_{\mu\nu}^a {\lambda^a\over 2}\,
(1+\gamma_5)b, \non
\en
with $q'=u,d,s$, $(\bar q_1q_2)_{_{V\pm A}}\equiv\bar
q_1\gamma_\mu(1\pm \gamma_5)q_2$, $O_3$--$O_6$ being the QCD
penguin operators, $O_{7}$--$O_{10}$ the electroweak penguin
operators, and $O_g$ the chromomagnetic dipole operator. After the
inclusion of vertex-type and penguin-type corrections, we obtain
\be \label{qcda}
 {\cal T} &=& {G_F\over\sqrt{2}}\Bigg\{~
V_{ub}V_{uq}^*\Big[\,a_1(\bar qu)_\vma\otimes(\bar
ub)_\vma+a_2(\bar uu)_\vma\otimes (\bar qb)_\vma\,\Big]   \non \\
&& \qquad+V_{cb}V_{cq}^*\Big[\,a_1(\bar qc)_\vma\otimes(\bar
cb)_\vma+a_2(\bar cc)_\vma\otimes (\bar qb)_\vma\,\Big]   \non \\
&& \qquad -V_{tb}V_{tq}^*\Big[\,a_3\sum _{q'}(\bar
q'q')_\vma\otimes (\bar qb)_\vma+a_4\sum_{q'}(\bar
qq')_\vma\otimes (\bar q' b)_\vma  \non \\ && \qquad\qquad\quad
+a_5\sum _{q'}(\bar q'q')_\vpa\otimes (\bar
qb)_\vma-2a_6\sum_{q'}(\bar qq')_\spp\otimes (\bar q' b)_\smp   \\
&& \qquad\qquad\quad +{3\over 2} a_7\sum _{q'}e_{q'}(\bar
q'q')_\vpa\otimes (\bar qb)_\vma-3a_8\sum_{q'}e_{q'}(\bar
qq')_\spp\otimes (\bar q' b)_\smp \non \\ && \qquad\qquad\quad
+{3\over 2} a_9\sum _{q'}e_{q'}(\bar q'q')_\vma\otimes (\bar
qb)_\vma+{3\over 2} a_{10}\sum_{q'}e_{q'}(\bar qq')_\vpa\otimes
(\bar q' b)_\vma\Big]\Bigg\},  \non
\en
where the symbol $\otimes$ stands for $\la MX|j_1\otimes j_2|B\ra=
\la M|j_1|0\ra\la X|j_2|B\ra$ or $\la X|j_1|0\ra\la M|j_2|B\ra$.

The coefficients in Eq. (\ref{qcda}) evaluated in the naive
dimensional regularization (NDR) scheme for $\gamma_5$ have the
expressions:
\be \label{ai}
 a_1 &=& c_1+{c_2\over N_c}+{\alpha_s\over 4\pi}\,{C_F\over N_c}c_2\,F,
\non \\
 a_2 &=& c_2+{c_1\over N_c}+{\alpha_s\over 4\pi}\,{C_F\over
N_c}c_1\,F, \non \\
 a_3 &=& c_3+{c_4\over N_c}+{\alpha_s\over
4\pi}\,{C_F\over N_c} c_4\,F, \non \\
 a_4 &=& c_4+{c_3\over
N_c}+{\alpha_s\over 4\pi}\,{C_F\over
N_c}\Bigg\{c_3\big[F+G(s_q)+G(s_b)+{4\over 3}
\big]-c_1\left({\lambda_u\over
\lambda_t}G(s_u)+{\lambda_c\over\lambda_t}G(s_c)-{2\over 3}\right)   \non \\
&& +(c_4+c_6) \sum_{i=u}^b G(s_i)  +{3\over 2} (c_8+c_{10})\sum
_{i=u}^b e_i G(s_i)-{1\over 2}c_9\big[G(s_q)+G(s_b)+{4\over 3}\big]+c_g G_g\Bigg\}, \non \\
 a_5 &=& c_5+{c_6\over
N_c}+{\alpha_s\over 4\pi}\,{C_F\over N_c} c_6(-F-12), \\
 a_6 &=& c_6+{c_5\over N_c}+{\alpha_s\over 4\pi}\,{C_F\over
N_c}\Bigg\{-c_1\left({\lambda_u\over
\lambda_t}G'(s_u)+{\lambda_c\over\lambda_t}G'(s_c)-{2\over 3}
\right)+c_3[G'(s_q)+G'(s_b)+{4\over 3}]-6c_5 \non \\ && +(c_4+c_6)
\sum_{i=u}^b G'(s_i) +{3\over 2} (c_8+c_{10})\sum _{i=u}^b e_i
G'(s_i)-{1\over 2}c_9\big[G'(s_q)+G'(s_b)+{4\over 3}\big]+c_g
G'_g\Bigg\}, \non
\\
 a_7 &=& c_7+{c_8\over N_c}+{\alpha_s\over 4\pi}\,{C_F\over N_c}
c_8(-F-12)-{\alpha\over 9\pi}\,N_cC_e,  \non \\
 a_8 &=& c_8+{c_7\over N_c}-6\,{\alpha_s\over 4\pi}\,{C_F\over
N_c}c_7-6{\alpha\over 9\pi}\,C'_e,  \non \\ a_9 &=&
c_9+{c_{10}\over N_c}+{\alpha_s\over 4\pi}\,{C_F\over N_c}
c_{10}\,F-{\alpha\over 9\pi}\,N_cC_e,  \non \\
 a_{10} &=&
c_{10}+{c_9\over N_c}+{\alpha_s\over 4\pi}\,{C_F\over N_c}
c_9\,F-{\alpha\over 9\pi}\,C_e, \non
\en
where $C_F=(N_c^2-1)/(2N_c)$, $s_i=m_i^2/m_b^2$, and $\lambda_{q}=
V_{qb}V^*_{qq'}$.

In Eq. (\ref{ai}), the vertex correction in the NDR scheme is
given by \cite{BBNS1} \be F=-12\ln{\mu\over m_b}-18+f_I, \en with
\be f_I=\int^1_0dx\,\Phi^M(x)\left(3{1-2x\over 1-x}\ln
x-3i\pi\right) \label{fI}, \en where $\Phi^M(x)$ is the leading
twist light-cone distribution amplitude (LCDA) of the light meson
$M$. For the vector meson $V$, $\Phi^V(x)$ is dominated by the
longitudinal DA ($\Phi^V_\|(x)$) as the contribution from the
transverse LCDA ($\Phi^V_\perp(x)$) is suppressed by a factor of
$m_V/m_b$. If the emitted meson is the $\J$ meson, then one has to
take into account the charmed quark mass effect \cite{CYJ,Chay}:
\be \label{fIJ}
 f_I^\J &=& \int^1_0 dx\,\Phi^\J_\|(x)\Bigg\{
3{1-2x\over 1-x}\ln x-3i\pi+3\ln(1-z)+2{z(1-x)\over 1-zx}   \non
\\ &+&\left({1-x\over (1-zx)^2}-{x\over
[1-z(1-x)]^2}\right)z^2x\ln zx+{z^2x^2[\ln(1-z)-i\pi]\over
[1-z(1-x)]^2}   \non \\ &+& 4rz\left[ \left({1\over
1-z(1-x)}-{1\over 1-zx}\right)\ln zx-{\ln(1-z)-i\pi\over
1-z(1-x)}\right]\Bigg\},
 \en
where $z=m_\J^2/m_B^2$ and $r=(f_\J^T m_c)/(f_\J m_\J)$, with
$f_\J^T$ being the tensor decay constant of the $\J$ defined by
 \be
 \la\J(P,\lambda)|\bar c\sigma_{\mu\nu}c|0\ra = -if_\J^T
(\vp^{*(\lambda)}_\mu P_\nu-\vp^{*(\lambda)}_\nu P_\mu).
 \en
Note that the third line in Eq. (\ref{fIJ}) arises from the
transverse polarization component of the $\J$. Since the
asymptotic form of the distribution amplitudes $\Phi^\J_\perp(x)$
and $\Phi^\J_\|(x)$ is the same, namely $6x(1-x)$, we will thus
assume $\Phi^\J_\perp(x)=\Phi^\J_\|(x)$ in general. It should be
stressed that the strong phase in $f_I$ or $f_I^\J$ comes from
final-state rescattering due to the hard gluon exchange between
the outgoing $M$ and $q$.

There are QCD penguin-type diagrams induced by the 4-quark
operators $O_i$ for $i=1,3,4,6,8,9,10$. The corrections are
described by the penguin-loop functions $G(s)$ and
$G'(s)=G_p(s)+G_\sigma(s)$ given by
 \be
 G(s) &=& -{4\over 3}\ln{\mu\over m_b}+4\int^1_0 dx\,\Phi^{M}(x)\int^1_0
du\,u(1-u)\ln[s-x u(1-u)], \non
\\
G_p(s) &=& -\ln{\mu\over m_b}+3\int^1_0 dx\,\Phi^{P}_p(x)\int^1_0
du\,u(1-u)\ln[s-x u(1-u)], \non
\\ G_\sigma(s) &=& -{1\over 3}\ln{\mu\over m_b}+{1\over 3}\int^1_0
dx\,{\Phi^{P}_\sigma(x)\over x}\int^1_0 du\,u(1-u)  \non \\
&\times& \Big\{\ln[s-x u(1-u)]-{1\over 2}\,{x u(1-u)\over s-x
u(1-u)}\Big\},  \label{G}
 \en
 where $\Phi^P_p$ and
$\Phi^P_\sigma$ are the twist-3 LCDAs of a pseudoscalar meson $P$,
to which we will come back shortly. It should be stressed that the
penguin-loop contribution proportional to $G'(s)$ is available
only if the emitted meson is of the pseudoscalar type. In Eq.
(\ref{ai}) we have also included the leading electroweak
penguin-type diagrams induced by the operators $O_1$ and $O_2$
\cite{Ali}: \be C_e &=& \left({\lambda_u\over
\lambda_t}G(s_u)+{\lambda_c\over
\lambda_t}G(s_c)-{2\over 3}\right)\left(c_2+{c_1\over N_c}\right), \non \\
C'_e &=& \left({\lambda_u\over \lambda_t}G'(s_u)+{\lambda_c\over
\lambda_t}G'(s_c)-{2\over 3}\right)\left(c_2+{c_1\over
N_c}\right). \en The dipole operator $O_g$ will give a tree-level
contribution proportional to \be G_g &=&
-2\int^1_0dx\,{\Phi^{M}(x)\over x}, \non
\\ G'_g &=& -{3\over 2}\int^1_0dx\,\Phi^{M}_p(x)-{1\over 6}\int^1_0
dx\,{\Phi^M_\sigma(x)\over x}.
\en

It is well known \cite{BSS} that strong phases can be
perturbatively generated from the penguin loop functions $G(x)$
and  $G'(x)$.
The virtual gluon's momentum squared $k^2=(1-x)m_B^2$ is no longer
an arbitrary parameter; it is convoluted with the emitted meson
wave function $\Phi^M(x)$.

The twist-3 DAs $\phi^P_p$ and $\phi^P_\sigma$ are defined in the
pseudoscalar and tensor matrix elements \cite{Ballp}:
\be
\la P(p)|\bar q_1(0)i\gamma_5 q_2(x)|0\ra &=& f_P
\mu_\chi^P\int^1_0 d\deta\,e^{i\deta p\cdot x}\phi_p^P(\deta),
\non
\\ \la P(p)|\bar q_1(0)\sigma_{\mu\nu}\gamma_5 q_2(x)|0\ra &=&
-{i\over 6}f_P\mu_\chi^P\left[1-\left({m_1+m_2\over
m_P}\right)^2\right] \non \\ && \times(p_\mu x_\nu-p_\nu x_\mu)
\int^1_0 d\deta \,e^{i\deta p\cdot x}\phi_\sigma^P(\bar \eta),
\label{tw3}
\en
where $\mu_\chi^P=m_P^2/(m_1+m_2)$ is proportional to the chiral
condensate. In the present paper we will take the asymptotic forms
for twist-2 and twist-3 distribution amplitudes:
\be
\Phi^{P,V,\J}(x)=6x(1-x), \qquad \Phi^P_p(x)=1, \qquad
\Phi^P_\sigma(x)=6x(1-x). \label{LCDA}
\en

Several remarks are in order. (i) The coefficients $a_{8a}$ and
$a_{10a}$ \cite{Muta} induced by the electroweak penguin operators
$O_{8,9,10}$ are absorbed in our case  into $a_6$ and $a_4$ in Eq.
(\ref{ai}). The contributions of $C_e$ and $C'_e$ to the
electroweak coefficients $a_{7-10}$, which have been neglected in
most recent literature, are numerically more important than
$a_{8a}$ and $a_{10a}$ owing to the large Wilson coefficients
$c_1$ and $c_2$. (ii) The scale and $\gamma_5$-scheme dependence
of the Wilson coefficients $c_i(\mu)$ is canceled by the
perturbative radiative corrections. In particular it is easily
seen that the scale dependence of $a_i$ is compensated by the
logarithmic $\mu$ dependence in $F$. However, note that the
coefficients $a_6$ and $a_8$ are scale dependent. This is because
the hadronic matrix element of $(S-P)(S+P)$ operator is
proportional to $\mu_\chi^P/m_b$ which is also scale dependent
owing to the running quark masses, the scale dependence of $a_6$
and $a_8$ is canceled by the corresponding one in
$\mu_\chi(\mu)/m_b(\mu)$. We have included penguin-type
corrections to $a_{4,6,8}$. Moreover, we have included the new
contributions from the twist-3 DA $\Phi_\sigma(x)$. (iii) In the
generalized factorization approach for nonleptonic decays, the
nonfactorized effects are sometimes parametrized in terms of the
effective number of colors $\nc$ so that \be a_{2i}=c_{2i}+{1\over
(\nc)_{2i}}c_{2i-1},\quad\qquad a_{2i-1}=c_{2i-1}+{1\over
(\nc)_{2i-1}}c_{2i}. \label{effai} \en Furthermore, it is assumed
that $(\nc)_i$ is process independent. In the improved QCD
factorization approach, $a_i$ (see Table II) and hence $(\nc)_i$
are in general complex and they are process and $i$ dependent.

\section{Two-body decays of the $\lowercase{b}$ quark}
The general decay amplitudes for $b\to Pq$ and $b\to Vq$ read
\be
A(b\to Pq) &=& i{G_F\over\sqrt{2}}\Big\{
(A^tV_{ub}V_{uq'}^*-A^pV_{tb}V_{tq'}^*)f_P p_P^\mu\,\bar
q\gamma_\mu(1-\gamma_5)b \non \\ &&- B\,V_{tb}V_{tq'}^*f_P\, \bar
q(1-\gamma_5)b\Big\}, \non \\ A(b\to Pc) &=& i{G_F\over\sqrt{2}}\,
A^tV_{cb}V_{uq'}^*f_P p_P^\mu\,\bar c\gamma_\mu(1-\gamma_5)b, \non
\\ A(b\to Vq) &=& {G_F\over\sqrt{2}} (\tilde
A^tV_{ub}V_{uq'}^*-\tilde A^p V_{tb}V_{tq'}^*)f_V
m_V\varepsilon_V^{*\mu}\,\bar q\gamma_\mu(1-\gamma_5)b,  \non \\
A(b\to Vc) &=& {G_F\over\sqrt{2}}\, \tilde A^tV_{cb}V_{uq'}^*f_V
m_V\varepsilon_V^{*\mu}\,\bar q\gamma_\mu(1-\gamma_5)b,  \non \\
A(b\to \J\,q) &=& {G_F\over\sqrt{2}} (\tilde
A^tV_{cb}V_{cq}^*-\tilde A^p V_{tb}V_{tq}^*)f_\J
m_\J\varepsilon_\J^{*\mu}\,\bar q\gamma_\mu(1-\gamma_5)b,
\label{bMq}
 \en
where $q'=d,s$ is not necessarily the same as $q$, and the
superscripts $t$ and $p$ denote tree and penguin
contributions, respectively. The coefficients $A$ and $B$ relevant
for some two-body hadronic $b$ decay modes of interest are
summarized in Table I. Owing to the complications for the $\eta$
and $\eta'$ production, their cases will be discussed separately
below. Note that the coefficient $B$ proportional to
$\mu_\chi/m_b$ is formally power suppressed in the heavy quark
limit, but numerically it is important since $\mu_\chi/ m_b\sim
12\Lqcd/m_b$ [see a discussion after Eq. (\ref{chi})]. Therefore,
we will keep this calculable power correction.

\begin{table}[ht]
\caption{The coefficients $A^t(\tilde A^t)$, $A^p(\tilde A^p)$ and
$B$ (in units of $\mu_\chi/m_b$) defined in Eq. (\ref{bMq}) for
some modes of interest.}
\begin{center}
\begin{tabular}{ l c c c c }
 Mode & $q'$ & $A^t(\tilde A^t)$ & $A^p(\tilde A^p)$ & $B(\mu_\chi/m_b)$
\\ \hline
 $b\to \pi^-u$ & $d$ & $a_1$ & $a_4+a_{10}$ & $2(a_6+a_8)$ \\
 $b\to \rho^-u$ & $d$ & $a_1$ & $a_4+a_{10}$ & \\
 $b\to \pi^0d$ & $d$ & $a_2/\sqrt{2}$ & $(-a_4-{3\over 2}a_7+{3\over 2}a_9+{1\over
 2}a_{10})/\sqrt{2}$ & $3a_8/\sqrt{2}$ \\
 $b\to \rho^0d$ & $d$ & $a_2/\sqrt{2}$ & $(-a_4+{3\over 2}a_7+{3\over 2}a_9+{1\over
 2}a_{10})/\sqrt{2}$ & \\
 $b\to \omega\,d$ & $d$ & $a_2/\sqrt{2}$ & $[2a_3+a_4+2a_5+{1\over 2}
 (a_7+a_9-a_{10})]/\sqrt{2}$ & \\
 $b\to\phi\,d$ & $d$ &  & $a_3+a_5-{1\over 2}(a_7+a_9)$ & \\
 $b\to\eta\,d$ & $d$ & & see~text & \\
 $b\to\eta^{'}d$ & $d$ & & see~text & \\
 $b\to \pi^- c$ & $d$ & $a_1$ & & \\
 $b\to \rho^- c$ & $d$ & $a_1$ & & \\
 $b\to K^0s$ & $d$ & & $a_4-{1\over 2} a_{10}$ & $2a_6-a_8$ \\
 $b\to K^{*0}s$ & $d$ & & $a_4-{1\over 2}a_{10}$ &  \\
 $b\to K^-u$ & $s$ & $a_1$ & $a_4+ a_{10}$ & $2(a_6+a_8)$ \\
 $b\to K^{*-}u$ & $s$ & $a_1$ & $a_4+ a_{10}$ & \\
 $b\to \bar K^0d$ & $s$ & & $a_4-{1\over 2} a_{10}$ & $2a_6-a_8$ \\
 $b\to \bar K^{*0}d$ & $s$ & & $a_4-{1\over 2}a_{10}$ &  \\
 $b\to K^-c$ & $s$ & $a_1$ & & \\
 $b\to K^{*-}c$ & $s$ & $a_1$ & & \\
 $b\to\eta\,s$ & $s$ & & see~text & \\
 $b\to\eta^{'}s$ & $s$ & & see~text & \\
 $b\to\pi^0s$ & $s$ & $a_2/\sqrt{2}$ & ${3\over 2\sqrt{2}}(-a_7+a_9)$ & \\
 $b\to\rho^0s$ & $s$ & $a_2/\sqrt{2}$ & ${3\over 2\sqrt{2}}(a_7+a_9)$ & \\
 $b\to\omega\,s$ & $s$ & $a_2/\sqrt{2}$ & $[2a_3+2a_5+{1\over 2}
 (a_7+a_9)]/\sqrt{2}$ & \\
 $b\to\phi\,s$ & $s$ & & $a_3+a_4+a_5-{1\over 2}(a_7+a_9+a_{10})$ & \\
 $b\to\J\, s$ & $s$ & $a_2$ & $a_3+a_5+a_7+a_9$ & \\
 $b\to\J\, d$ & $d$ & $a_2$ & $a_3+a_5+a_7+a_9$ & \\
\end{tabular}
\end{center}
\end{table}

The decay amplitudes of $b\to \eta^{(')}s,~\eta^{(')}d$ have the
expressions
\be
A(b\to \eta^{(')}s) &=& i{G_F\over\sqrt{2}}\Bigg\{\Bigg[
V_{ub}V_{us}^*a_2f_\etap^u+V_{cb}V_{cs}^*a_2f_\etap^c
-V_{tb}V_{ts}^*\Big((a_3-a_5-a_7+a_9)f_\etap^c \non
\\ &&  + [a_3+a_4-a_5+{1\over
2}(a_7-a_9-a_{10})]f_\etap^s \non \\ &&+(2a_3-2a_5-{1\over
2}a_7+{1\over 2}a_9)f_\etap^u\Big)\Bigg]\,p_\etap^\mu\bar
s\gamma_\mu(1-\gamma_5)b \non
\\ &&+(2a_6-a_8){m^2_\etap\over
2m_s(m_b-m_s)}(f_\etap^s-f_\etap^u) \bar s(1-\gamma_5)b\Bigg\},
\label{etaps}
\en
and
\be
A(b\to \eta^{(')}d) &=& i{G_F\over\sqrt{2}}\Bigg\{\Bigg[
V_{ub}V_{ud}^*a_2f_\etap^u+V_{cb}V_{cd}^*a_2f_\etap^c
-V_{tb}V_{td}^*\Big((a_3-a_5-a_7+a_9)f_\etap^c \non
\\ && +[2a_3+a_4-2a_5+{1\over
2}(-a_7+a_9-a_{10})]f_\etap^u \non \\ && + [a_3-a_5+{1\over
2}(a_7-a_9)]f_\etap^s\Big) \Bigg]\,p_\etap^\mu\bar
d\gamma_\mu(1-\gamma_5)b \non
\\ &&+(2a_6-a_8){m^2_\etap\over
2m_s(m_b-m_d)}(f_\etap^s-f_\etap^u)r_\etap \bar
d(1-\gamma_5)b\Bigg\}, \label{etapd}
\en
where the decay constants of the $\eta$ and $\eta'$ are defined by
$\la 0|\bar q\gamma_\mu\gamma_5 q|\etap\ra=if^q_\etap p_\mu$. The
$\eta'$ decay constants follow a two-angle mixing pattern
\cite{Leutwyler,Kroll}:
\be
f_{\eta'}^u={f_8\over\sqrt{6}}\sin\theta_8+{f_0\over\sqrt{3}}\cos\theta_0,
\qquad
f_{\eta'}^s=-2{f_8\over\sqrt{6}}\sin\theta_8+{f_0\over\sqrt{3}}\cos
\theta_0,
\en
with $f_8$ and $f_0$ being the decay constants of the SU(3) octet
and singlet $\eta_8$ and $\eta_0$, respectively:
\be
\la 0|A_\mu^0|\eta_0\ra=if_0 p_\mu, \qquad \la 0|A_\mu^8|\eta_8
\ra=if_8 p_\mu.
\en
Likewise, for the $\eta$ meson
\be
f_{\eta}^u={f_8\over\sqrt{6}}\cos\theta_8-{f_0\over\sqrt{3}}\sin\theta_0,
\qquad
f_{\eta}^s=-2{f_8\over\sqrt{6}}\cos\theta_8-{f_0\over\sqrt{3}}\sin
\theta_0.
\en
It must be emphasized that the two-mixing angle formalism
proposed in \cite{Leutwyler,Kroll} applies to the decay constants
of the $\eta'$ and $\eta$ rather than to their wave functions. It
is found in \cite{Kroll} that phenomenologically
\be
\label{theta} \theta_8=-21.2^\circ, \qquad \theta_0=-9.2^\circ,
\qquad f_8/f_\pi=1.26,  \qquad f_0/f_\pi=1.17.
\en
Numerically, we obtain
\be   \label{fdecay}
 f_{\eta}^{u,d}=78\,{\rm MeV}, \quad f_\eta^s=-112\,{\rm MeV},
 \quad f_{\eta'}^{u,d}= 63\,{\rm
MeV}, \quad f_{\eta'}^s=137\,{\rm MeV}.
\en

The decay constant $f_{\eta'}^c$, defined by $\la 0|\bar
c\gamma_\mu\gamma_5c|\eta'\ra=if_{\eta'}^c q_\mu$, is related to
the intrinsic charm content of the $\eta'$ and it has been
estimated from theoretical calculations \cite{Halperin} and from
the phenomenological analysis of the data on
$J/\psi\to\eta_c\gamma,\,J/\psi \to\eta'\gamma$ and of the
$\eta\gamma$ and $\eta'\gamma$ transition form factors
\cite{Ali,Kroll,Petrov}; it is expected to lie in the range --2.0 MeV $\leq
f_{\eta'}^c\leq$ --18.4 MeV. In this paper we use the values
\be
f_{\eta'}^c=-(6.3\pm 0.6)\,{\rm MeV},  \qquad f_\eta^c=-(2.4\pm
0.2) \,{\rm MeV},
\en
as obtained from a phenomenological analysis performed in
\cite{Kroll}.

Care must be taken to consider the pseudoscalar matrix element for
$\etap\to$ vacuum transition. The anomaly effects must be included
in order to ensure a correct chiral behavior for the pseudoscalar
matrix element \cite{CT98}. The results are \cite{Kagan,Ali}
\be
 \label{anomaly} \la\etap|\bar s\gamma_5 s|0\ra &=&
-i{m_{\etap}^2\over 2m_s}\,\left(f_{ \etap}^s-f^u_{\etap}\right),
\non \\ \la\etap|\bar u\gamma_5u|0\ra &=& \la\etap|\bar
d\gamma_5d|0\ra=r_{\etap} \,\la\etap|\bar s\gamma_5s|0\ra,
\en
with \cite{CT98}
\be \label{reta}
 r_{\eta'} &=& {\sqrt{2f_0^2-f_8^2}\over\sqrt{2f_8^2-f_0^2}}\,{\cos\theta+
{1\over \sqrt{2}}\sin\theta\over \cos\theta-\sqrt{2}\sin\theta},
\non \\ r_{\eta} &=& -{1\over
2}\,{\sqrt{2f_0^2-f_8^2}\over\sqrt{2f_8^2-f_0^2}}\,
{\cos\theta-\sqrt{2}\sin\theta\over \cos\theta+{1\over\sqrt{2}}
\sin\theta},
\en
where $\theta$ is the $\eta-\eta'$ mixing angle:
\be
\eta'=\eta_8\sin\theta+\eta_0\cos\theta, \qquad
\eta=\eta_8\cos\theta-\eta_0 \sin\theta.
\en
We will use $\theta=-15.4^\circ$ as determined in \cite{Kroll}.

To proceed with numerical calculations, we need to specify some input
parameters. For Wilson coefficients, we use the next-to-leading
ones in the NDR scheme \be c_1=1.082, \quad c_2=-0.185,\quad
c_3=0.014, \quad c_4=-0.035, \quad c_5=0.009, \quad c_6=-0.041,
\non \\ c_7/\alpha=-0.002, \quad c_8/\alpha=0.054, \quad
c_9/\alpha=-1.292, \quad c_{10}/\alpha=0.263,\quad c_g=-0.143, \en
with $\alpha$ being the electromagnetic fine-structure coupling
constant. These values taken from Table XXII of \cite{Buras96} are
evaluated at $\mu=\ov m_b(m_b)=4.40$ GeV and
$\Lambda^{(5)}_{\ov{\rm MS}}=225$ MeV. For the decay constants
other than $\eta$ and $\eta'$ we employ \be &&f_\pi=132\,{\rm
MeV},\qquad f_K=160\,{\rm MeV}, \qquad f_\rho=216\,{\rm
MeV},\qquad f_{K^*}=221\,{\rm MeV}, \non \\ && f_\omega=195\,{\rm
MeV},\qquad f_\phi=237\,{\rm MeV}, \qquad f_\J=405\,{\rm MeV}. \en
For the CKM matrix elements, we take $|V_{cb}|=0.0395$ and
$|V_{ub}/V_{cb}|=0.085$. For the unitarity angle $\gamma$ we use
$\gamma=60^\circ$ as extracted from recent global CKM fits
\cite{Faccioli}. The other two unitarity angles $\alpha$
and $\beta$ are then fixed.

The hadronic matrix element of the $(S-P)(S+P)$ operator involves
the quantity $\mu_\chi^P/\ov m_b$ \be {\mu_\chi^P\over \ov
m_b(\mu)}={m_P^2\over \ov m_b(\mu)[\ov m_1(\mu)+\ov m_2(\mu)]},
\label{chi} \en where $\ov m_q$ is the running quark mass of the
quark $q$, and hence it is formally $\Lqcd/\ov m_b$ power
suppressed in the heavy quark limit. However, numerically it is
chirally enhanced. At $\mu=m_b$ scale, we have (see \cite{Koide}
for the running quark mass ratios and evolution) \be \ov
m_s(m_b)=90\,{\rm MeV},\qquad \ov m_d(m_b)=4.6\,{\rm MeV},\qquad
\ov m_u(m_b)=2.3\,{\rm MeV}, \en corresponding to $\ov m_s(1\,{\rm
GeV})=140\,{\rm MeV}$ or $\ov m_s(2\,{\rm GeV})=101\,{\rm MeV}$.
We will neglect the small flavor-SU(3) breaking for the chiral
condensate and use the averaged value $\mu_\chi(m_b)=2.7$
GeV.\footnote{A value as small as $\mu_\chi=1.4$ GeV is sometimes
chosen in the literature. However, we favor the higher value given
above as one can then readily account for
the observed $B\to K\pi$ rates, which are difficult to explain in
terms of the smaller $\mu_\chi$.} As a consequence, $\mu_\chi/\ov
m_b=12\Lqcd/\ov m_b$ yields a large chiral enhancement. Apart from
the current quark masses appearing in the use of equations of
motion, we will utilize the pole masses for the heavy quarks:
$m_b=4.8$ GeV and $m_c=1.5$ GeV. To compute the branching ratios,
we use the mean lifetime of the admixture of bottom particles:
$\tau=1.564\times 10^{-12}s$ \cite{PDG}.

\begin{table}[ht]
\caption{\small Numerical values for the coefficients $a_i$ (in
units of $10^{-4}$ for $a_3,\cdots,a_{10}$) in QCD factorization
and in naive factorization for $b\to Pq$ and $q'=d,s$ .}
\begin{center}
\begin{tabular}{ c c c c }
 & \multicolumn{2}{c}{QCD factorization} & \\\cline{2-3}
\raisebox{1.5ex}[0cm][0cm]{} &  $q'=d$ & $q'=s$ &
\raisebox{1.5ex}[0cm][0cm]{naive factorization} \\ \hline
 $a_1$ & $1.05+0.01i$ & $1.05+0.01i$ & 1.02 \\
 $a_2$ & $0.02-0.08i$ & $0.02-0.08i$ & 0.18 \\
 $a_3$ & $74+26i$ & $74+26i$ & 23 \\
 $a_4$ & $-317-29i$ & $-353-58i$ & $-303$ \\
 $a_5$ & $-67-30i$ & $-67-30i$ & $-8$ \\
 $a_6$ & $-400-27i$ & $-440-45i$ & $-380$ \\
 $a_7$ & $-0.29-0.65i$ & $-0.89-1.13i$ & 0.80 \\
 $a_8$ & $3.43-0.36i$ & $3.22-0.46i$ & 3.89 \\
 $a_9$ & $-92.3-2.3i$ & $-92.9-2.8i$ & $-87.9$ \\
 $a_{10}$ & $0.8+6.6i$ & $0.6+6.4i$ & $-12.2$ \\
\end{tabular}
\end{center}
\end{table}

Since the coefficients $a_{4,6-10}$ involve the CKM matrix
elements $\lambda_{u,c}/\lambda_t$ [see Eq. (\ref{ai})], the
results of the coefficients $a_i$  are exhibited in Table II  for
$b\to P\,q$ and $q'=d,s$. It is evident that $a_{3,4,5,6}$ are
enhanced substantially compared to the naive factorization values
and have large imaginary parts and that $a_2$ in QCD factorization
becomes very small. In particular, the calculated coefficient
$a_2(Pq)=0.02-0.08i$ in QCD factorization is dramatically
different from the value 0.18 obtained in naive factorization. The
smallness of $a_2$ imposes a serious problem. For example, the
predicted branching ratio of $b\to\J s$ is too small compared to
the experimental value ${\cal B}(B\to \J X_s)=(8.0\pm 0.8)\times
10^{-3}$ \cite{JX}. We will come to this point in Sec. IV. Note
that $a_9$ is the dominant electroweak penguin coefficient. Owing
to the large cancellation between $a_3$ and $a_5$, the decay
$b\to\phi\,d$ is electroweak penguin dominated. Likewise,
$b\to\pi^0s$ and $\rho^0s$ are also dominated by electroweak
penguin diagrams.

We are now ready to compute the decay rates using,
\be
\Gamma(b\to Pq) &=& {G^2_Ff_P^2m_b^3\over
16\pi}\Big(|A^tV_{ub}V_{uq'}^*-A^pV_{tb}V_{tq'}^*|^2
+|B\,V_{tb}V_{tq'}^*|^2\Big)\left(1-{m_P^2\over m_b^2}\right),
\non \\ \Gamma(b\to Pc) &=& {G^2_Ff_P^2m_b^3\over
16\pi}\,|A^tV_{cb}V_{uq'}^*|^2\left(1+{m_c^2\over
m_b^2}-{m_P^2\over m_b^2}\right), \non
\\ \Gamma(b\to Vq) &=& {G^2_Ff_V^2m_b^3\over 16\pi}|\tilde
A^tV_{ub}V_{uq'}^*-\tilde A^pV_{tb}V_{tq'}^*|^2\left(1+{m_V^2\over
m_b^2}-2{m_V^4\over m_b^4}\right), \non \\ \Gamma(b\to Vc) &=&
{G^2_Ff_V^2m_b^3\over 16\pi}|\tilde
A^tV_{cb}V_{uq'}^*|^2\left(1-{m_c^2-m_V^2\over
m_b^2}+{m_c^4-2m_V^4\over m_b^4}\right), \non
\\\Gamma(b\to \J\,q) &=& {G^2_Ff_\J^2m_b^3\over 16\pi}|\tilde
A^tV_{cb}V_{cq}^*-\tilde A^pV_{tb}V_{tq}^*|^2\left(1+{m_\J^2\over
m_b^2}-2{m_\J^4\over m_b^4}\right).
\en
The expression of $\Gamma(b\to\etap q)$ is similar to that of
$\Gamma(b\to Pq)$. We will also study the $CP$-violating
partial-rate asymmetry (PRA) defined by
\be
{\cal A}=\,{\Gamma(b\to Mq)-\Gamma(\bar b\to \bar M\bar q)\over
\Gamma(b\to Mq)+\Gamma(\bar b\to \bar M\bar q)}.
\en

\begin{table}[ht]
\caption{$CP$-averaged branching ratios and partial-rate
asymmetries for some two-body hadronic $b$ decays. For comparison,
the predicted branching ratios and rate asymmetries (in absolute
values for $\gamma=60^\circ$) based on naive factorization [4] are
given in parentheses.}
\begin{center}
\begin{tabular}{ l l c  }
 Mode & BR & PRA(\%)  \\ \hline
 $b\to \pi^-u$ & $1.5\times 10^{-4}~(1.3\times 10^{-4})$ & -2~(7) \\
 $b\to \rho^-u$ & $4.2\times 10^{-4}~(3.5\times 10^{-4})$ & -2~(7)  \\
 $b\to \pi^0d$ & $5.3\times 10^{-7}~(2.4\times 10^{-6})$ & 93~(31)  \\
 $b\to \rho^0d$ &  $1.4\times 10^{-6}~(5.9\times 10^{-6})$ & 91~(33)  \\
 $b\to \omega\,d$ & $2.5\times 10^{-6}~(5.8\times 10^{-6})$ & -97~(34)  \\
 $b\to\phi\,d$ & $6.9\times 10^{-8}~(2.3\times 10^{-7})$ & -2~(0)  \\
 $b\to\pi^- c$ & $2.2\times 10^{-2}$ & 0 \\
 $b\to\rho^- c$ & $5.1\times 10^{-2}$ & 0 \\
 $b\to\eta\,d$ & $1.5\times 10^{-6}$ & -59 \\
 $b\to\eta' d$ & $1.0\times 10^{-6}$ & 38 \\
 $b\to K^0s$ & $4.0\times 10^{-6}~(2.5\times 10^{-6})$ & -20~(4)   \\
 $b\to K^{*0}s$ & $2.6\times 10^{-6}~(2.9\times 10^{-6})$ & -24~(14)   \\
 $b\to K^-u$ & $9.2\times 10^{-5}~(2.9\times 10^{-5})$ & 5~(28)  \\
 $b\to K^{*-}u$ & $4.8\times 10^{-5}~(5.1\times 10^{-5})$ & 17~(44)  \\
 $b\to \bar K^0d$ & $1.0\times 10^{-4}~(2.0\times 10^{-5})$ & 0.8~(1)  \\
 $b\to \bar K^{*0}d$ & $6.6\times 10^{-5}~(2.6\times 10^{-5})$ & 0.9~(3)   \\
 $b\to K^- c$ & $1.7\times 10^{-3}$ & 0 \\
 $b\to K^{*-}c$ & $2.7\times 10^{-3}$ & 0 \\
 $b\to\eta\,s$ & $1.9\times 10^{-5}$ & -4 \\
 $b\to\eta' s$ & $5.4\times 10^{-5}$ & 1 \\
 $b\to\pi^0s$ & $1.8\times 10^{-6}~(1.6\times 10^{-6})$ & 19~(0)  \\
 $b\to\rho^0s$ & $5.1\times 10^{-6}~(4.3\times 10^{-6})$ & 19~(0)  \\
 $b\to\omega\,s$ & $3.3\times 10^{-7}~(1.3\times 10^{-6})$ & 61~(0)  \\
 $b\to\phi\,s$ & $5.5\times 10^{-5}~(6.3\times 10^{-5})$ & 1~(0)  \\
 $b\to\J\, s$ & $5.4\times 10^{-4}$ & -0.5  \\
 $b\to\J\, d$ & $2.8\times 10^{-5}$ & 10  \\
\end{tabular}
\end{center}
\end{table}

The $CP$-averaged branching ratios and direct $CP$-violating
partial rate asymmetries for some two-body hadronic $b$ decays of
interest are summarized in Table III. Compared to the predictions
of branching ratios based on naive factorization \cite{Soni},
there are three major modifications: (i) Decay modes $\pi^- u$,
$\bar K^0d$, $\bar K^{*0}d$ and $K^-u$ are significantly enhanced
owing to the large penguin coefficients $a_6$ and $a_4$. (ii) The
modes $\pi^0d,~\rho^0d,~\omega\,d,~\J s,~\J d$ with neutral
emitted mesons are suppressed relative to the naive factorization
ones due to the smallness of $a_2$. (iii) The $\phi d$ mode has a
smaller rate due to the large cancellation between $a_3$ and
$a_5$. That is, while $\phi d$ is QCD-penguin dominated in naive
factorization, it becomes electroweak-penguin dominated in QCD
factorization.

For the prompt $\eta'$ production in semi-inclusive decays, we
find the four-quark operator contributions to $b\to\eta's$ can
only account for about 10\% of the measured result \cite{etapK}:
\be
{\cal B}(B\to\eta' X_s)=(6.2\pm 1.6\pm 1.3^{+0.0}_{-1.5}({\rm
bkg}))\times 10^{-4} \qquad {\rm for}~2.0<p_{\eta'}<2.7~{\rm
GeV/c}, \label{etapX}
\en
where $X_s$ is the final state containing a strange quark. One
important reason is that there is an anomaly effect in the matrix
element $\la\eta'|\bar s\gamma_5s|0\ra$ manifested by the decay
constant $f_{\eta'} ^u$ [see Eq.~(\ref{anomaly})]. Since
$f_{\eta'}^u\sim {1\over 2}f_{\eta'}^s$ [cf. Eq.~(\ref{fdecay})],
it is obvious that the decay rate of $b\to\eta's$ induced by the
$(S-P)(S+P)$ penguin interaction is suppressed by the QCD anomaly
effect. If there were no QCD anomaly, one would have ${\cal
B}(b\to\eta's)=2.2\times 10^{-4}$ from four-quark operator
contributions which are about one third of the experimental value.

It is useful to explicitly take into account the constraints from
the CPT theorem when computing PRA's for inclusive decays at the
quark level \cite{Hou} (for a review, see \cite{Atwood}). Take
$b\to du\bar u$ as an example and note that the penguin amplitude,
say $\lambda_t a_4$, can be written as $-(\lambda_u
a_4^u+\lambda_c a_4^c)$, where $\lambda_q=V_{qb}V^*_{qq'}$,
$a_4^u=-c_1G(s_u)+\cdots$ and $a_4^c=-c_1G(s_c)+\cdots$ with the
ellipses being the common terms given in Eq. (\ref{ai}) for $a_4$
and $G(s_q)$ the penguin function with the internal quark $q$ [see
Eq. (\ref{G})]. In general, $a_4$ has absorptive contributions
from all $u,d,s,c$ quark loops. It is clear that PRA's arise from
the interference between the tree amplitude $\lambda_u a_1$ and
the penguin amplitude $\lambda_c a_4^c$ and the one between
$\lambda_u a_4^u$ and $\lambda_c a_4^c$. The CPT theorem implies
that the ``diagonal" strong penguin phases coming from the
diagonal process $q+\bar q \to q+\bar q$ will not contribute to
the rate asymmetry \cite{Wolf}. For example, at order $\alpha_s$
the interference between $\lambda_u a_1$ and $\lambda_c a_4^c$
with the absorptive cut from the $u$ quark loop in the penguin
diagram does not contribute to PRA in $b\to s(d)u\bar u$ decays.
It is easily seen that the compensating process for this
interference is itself. Likewise, the PRA at order $\alpha_s^2$
due to the interference between two different penguin amplitudes
with the $u\bar u$ absorptive cut in the penguin loop will be
compensated by the one between the tree amplitude and the higher
order penguin diagram that contains an absorptive part from the
$u\bar u$ quark loop inset in the gluon propagator. Therefore, one
has to disregard the penguin phases coming from $G(s_u)$ and
$G'(s_u)$ for the PRA's in the decay $b\to du\bar u$. Similarly,
the phase of the $s$-loop penguin diagram should be dropped in
rate-asymmetry calculations of $b\to ds\bar s$ and $b\to s s\bar
s$ in order to be consistent with the requirement of the CPT
theorem. By the same token, the strong ``diagonal" phase in
coefficients $a_i$ due to final-state hard gluon exchange [see Eq.
(\ref{fI})] will not contribute to rate asymmetries in quark level
processes.

The implication of the CPT theorem for PRA's at the hadron level
in exclusive or semi-inclusive reactions is more complicated
\cite{AtwoodCPT}.
Consider the above example $b\to du\bar u$ again. The
corresponding semi-inclusive decays of the $b$ quark can be
manifested as $b\to (\pi^-,\rho^-)u$ and $b\to
(\pi^0,\rho^0,\omega)d$ at the two-body level and $(\pi^-\pi^0,K^0
K^-)u$, $(\pi^+\pi^-,\pi^0\pi^0,K^+ K^-)d$ at the three-body level
and etc. The CPT theorem no longer constrains the absorptive cut
from the $u$-loop penguin diagram not to contribute separately to
each aforementioned semi-inclusive $b$ decay, though the
cancellation between $u\bar u$ and $c\bar c$ quarks will occur
when all semi-inclusive modes are summed over. In view of this
observation, we shall keep all the strong phases in the
calculation of direct $CP$ violation in the individual
semi-inclusive decay.

The presence of the strong phase in the hard kernel $f_I$ [Eq.
(\ref{fI})] in QCD factorization has several noticeable effects:
(i) The rate asymmetries in the decays $b\to \phi
d,~(\pi^0,\rho^0,\omega)s$ vanish in naive factorization because
the coefficients $a_2$ and $a_{3,5,7,9,10}$ there are real. In QCD
factorization, the large phase of $a_2$ (see Table II) will yield
large PRA's for these modes. (ii) The color-suppressed
tree-dominated decays $b\to(\pi^0,\rho^0,\omega)d$ have large
PRA's due to the large imaginary part of $a_2$  and the smallness
of $|a_2|$. To see this, we consider $b\to \rho^0 d$ as an
illustration. The partial rate difference $\Delta\Gamma(b\to\rho^0
d)=\Gamma(b\to \rho^0d)-\Gamma(\bar b\to\rho^0\bar d)$ is
proportional to (see Table I for the amplitude)
 \be
 \Delta\Gamma(b\to\rho^0 d)\propto
 {\rm Im}(V_{ub}V_{ud}^*V_{tb}^*V_{td})\,{\rm Im}\left[a_2(-a_4+{3\over 2}a_7+{3\over
 2}a_9+{1\over 2}a_{10})\right].
 \en
Since $a_2$ is dominated by the imaginary part, it follows that
 \be \label{CPtree}
\Delta \Gamma(b\to\rho^0 d)\propto
 \sin\alpha\, {\rm Im}a_2\,{\rm Re}\left(-a_4+{3\over 2}a_7+{3\over
 2}a_9+{1\over 2}a_{10}\right).
 \en
Because $|V_{ub}V_{ud}^*|\gg |V_{tb}V_{td}^*|$ and $|a_2|$ is
small, it is clear that $b\to\rho^0 d$ has a large PRA governed by
the vertex-induced strong phase. In contrast, the rate asymmetry
in penguin-dominated modes is largely due to the strong penguin
phase. Consider the process $b\to K^{*-}u$. The partial rate
difference is
 \be \label{CPpeng}
 \Delta\Gamma(b\to K^{*-}u)&\propto & {\rm Im}(V_{ub}V_{us}^*V_{cb}^*V_{cs}){\rm Im}
 [a_1a_4^c+a_4^ua_4^c]  \\ &\approx &  \sin\gamma\,[{\rm Im}a_1\,{\rm Re}
 a_4^c-c_1{\rm Im}G(s_c){\rm Re}a_1-c_1{\rm
 Im}G(s_u){\rm Re}a_4^c-c_1{\rm Im}G(s_c){\rm Re}a_4^u], \non
 \en
where we have applied the relation $\lambda_t a_4=-(\lambda_u
a_4^u+\lambda_c a_4^c)$ as given before. Since the imaginary part
of $a_1$ is very small (see Table II), it is evident that CP
asymmetries in the penguin-dominated modes are governed by the
strong penguin phase.

In the present QCD factorization approach we have considered the
penguin corrections to $a_{4,6-10}$ induced by not only tree
operators but also QCD and electroweak penguin operators. For
example, the parameters $a_{7,9,10}$ do receive an absorptive
contribution via electroweak penguin-type diagrams generated by
tree 4-quark operators $O_{1,2}$ [see Eq. (\ref{ai})]. Therefore,
they receive more penguin phases. Another important difference
between QCDF and naive factorization is that the gluon's
virtuality $k^2$ is no longer arbitrary; this tends to remove
considerable uncertainties in the estimates of the CP asymmetries.
These may account for the general differences between the present
results and \cite{Soni}. Note that our predictions for PRA's in
the decays $\bar B^0\to K^-(K^{*-})X$ and $B^-\to\bar K^0(\bar
K^{*0})X$ are in agreement with \cite{He} for $\gamma=60^\circ$.
(The definition of the rate asymmetry in \cite{He} has a sign
opposite to ours.)

\section{Semi-inclusive $B$ decays}
A major advantage of studying the quasi-two-body decay of the $b$
quark is that it does not involve the unknown form factors and
hence the theoretical uncertainty is considerably reduced. In
order to retain this merit for semi-inclusive $B$ decays, it is
necessary to circumvent the second matrix element term appearing
in Eq. (\ref{fm.e.}). Fortunately, this can be achieved by
choosing a $B$ meson whose spectator quark content is not
contained in the outgoing meson $M$. For example, the counterpart
of $b\to \pi^0 d$ at the meson level will be $\ov B^0_s\to \pi^0
X$ as then the  $\ov B_s-\pi^0$ transition does not contribute.
In contrast,
the decay $B^-\to \pi^0 X$ or $\ov B^0\to\pi^0 X$ will involve the
unwanted $B-\pi$ form factors. As stressed in Sec.II,
it is necessary to impose a
cut, say $E_M>2.1$ GeV for the light emitted meson,
in order to reduce contamination from the unwanted background and
enhance the presence of the two-body quark decay $b\to Mq$.
Therefore, in the absence of the bound state effect it is expected
that, for example, $\Gamma(\bar B^0\to \pi^-X)\approx \Gamma(b\to
\pi^- u)$ after applying the parton-model approximation
 \be
\sum_X|X\ra\la X|\approx \sum _s\int{d^3p\over
(2\pi)^32E_u}\,|u(p_u,s)\ra\la u(p_u,s)|.
 \en

There are two more complications for semi-inclusive $B$ decays.
First,  $B\to MX$ can be viewed as the two-body decay $b\to Mq$ in
the heavy quark limit. For the finite $b$ quark mass, it becomes
necessary to consider the initial $b$ quark bound state effect.
Second, consider the 3-body decay $\ov B\to Mq_1\bar q_2$ with the
quark content $(b\bar q_2)$ for the $\bar B$ meson. One needs a
hard gluon exchange between the spectator quark $\bar q_2$ and the
meson $M$ in order to ensure that the outgoing $\bar q_2$ is hard.
For exclusive two-body decays, the nonfactorizable hard spectator
contribution is customarily denoted as \cite{BBNS1}
\be
{G_F\over\sqrt{2}}\,{\alpha_s\over 4\pi}\,{C_F\over N_c}c_if_{II}.
\en
Numerically, $f_{II}$ is much larger than $f_I$ for exclusive
decays. For the semi-inclusive case at hand, it has been argued
that $f_{II}$ is subject to a phase-space suppression since it
involves three particles in the final states rather than the
two-body one for $f_I$ \cite{Hephi}. However, we shall see below
that it is not the case for color-suppressed decay modes.

\subsection{Initial bound state effect}
The initial bound state effects on branching ratios and $CP$
asymmetries have been studied recently in \cite{He} using two
different approaches: the light-cone expansion approach and the
heavy quark effective theory approach. We will follow \cite{He} to
employ the second approach which amounts to modifying the decay
rates by
\be
\Gamma(B\to PX) &=& {G_Ff_P^2m_b^3\over
16\pi}\Bigg\{|A^tV_{ub}V_{uq'}^*-A^pV_{tb}V_{tq'}^*|^2\eta_1
+|B\,V_{tb}V_{tq'}^*|^2\eta_2\Bigg\}\left( 1-{m_P^2\over
m_B^2}\right), \non
\\ \Gamma(B\to VX) &=& {G_Ff_V^2m_b^3\over 16\pi}|\tilde
A^tV_{ub}V_{uq'}^*-\tilde
A^pV_{tb}V_{tq'}^*|^2\eta_1\left(1+{m_V^2\over m_B^2}-2{m_V^4\over
m_B^4}\right),  \\ \Gamma(B\to \J\,X) &=& {G_Ff_\J^2m_b^3\over
16\pi}|\tilde A^tV_{cb}V_{cq}^*-\tilde
A^pV_{tb}V_{tq}^*|^2\eta_1\left(1+{m_\J^2\over
m_B^2}-2{m_\J^4\over m_B^4}\right), \non \en where \be
\eta_1=\left(1+{7\over 6}\,{\mu_G^2\over m_b^2}-{53\over
6}\,{\mu^2_\pi\over m_b^2}\right), \qquad
\eta_2=\left(1-{\mu_G^2\over 2m_b^2}+{\mu^2_\pi\over
2m_b^2}\right), \en and \be \mu_G^2={ \la B|\bar h
G_{\mu\nu}\sigma^{\mu\nu}h|B\ra\over 4m_B}, \qquad \mu_\pi^2=-{\la
B|\bar h(iD_\bot)^2h|B\ra\over 2m_B},
 \en
with $D_\bot^\mu=D^\mu-v^\mu v\cdot D$, where $v$ is the
four-velocity of the $B$ meson. The nonperturbative HQET parameter
$\mu_G^2$ is fixed from the $B^*-B$ mass splitting to be
$0.36\,{\rm GeV}^2$. Following \cite{He} we use
$\mu_\pi^2=0.5\,{\rm GeV}^2$, which is consistent with QCD sum
rule and lattice QCD calculations \cite{lambda2}. Compared to the
two-body decays $b\to Mq$ shown in Table III, we see that the
branching ratio of $B\to PX$ and $B\to VX$ owing to bound state
effects is reduced by a factor of $(5\sim 10)\%$ and $17\%$,
respectively, while the CP asymmetry remains intact for $VX$
decays and for most of $PX$ modes.

\subsection{Nonfactorizable hard spectator interactions}
We now turn to the hard spectator interactions in the 3-body decay
$B(p_B)\to M(p_M)+q_1(p_1)+\bar q_2(p_2)$ with a hard gluon
exchange between the spectator quark $\bar q_2$ and the meson $M$.
A straightforward calculation yields
\be
A_{\rm spect}(B\to Pq_1\bar q_2) & &
={G_F\over\sqrt{2}}\,{\alpha_s\over 4\pi}\,{C_F\over
N_c}\,{4\pi^2\over
N_c}\,f_Pf_B(A^t_{sp}V_{ub}V^*_{uq'}-A^p_{sp}V_{tb}V^*_{tq'}) \non
\\ && \times \int^1_0d\xi\,{\Phi^P(\xi)\over
\xi}\int^1_0d\drho\,{\Phi^B_1(\drho)\over\drho} \left(
{p_P^\mu\bar q_1\gamma_\mu(1-\gamma_5)q_2\over p_P\cdot
p_2}-{m_B\bar q_1(1+\gamma_5)q_2\over p_B\cdot p_2}\right), \non
\\
 A_{\rm spect}(B\to Vq_1\bar q_2) & &
=-i{G_F\over\sqrt{2}}\,{\alpha_s\over 4\pi}\,{C_F\over
N_c}\,{4\pi^2\over N_c}\,m_Vf_Vf_B(\tilde
A^t_{sp}V_{ub}V^*_{uq'}-\tilde A^p_{sp}V_{tb}V^*_{tq'})  \non \\
&& \times \int^1_0d\xi\int^1_0d\drho\,
{\Phi^V(\xi)\Phi^B_1(\drho)\over \xi\,\drho\,(p_B\cdot p_2)
(p_V\cdot p_2+\xi m_V^2/2)} \non \\ && \times \Bigg\{  (p_B\cdot
p_2-\xi p_B\cdot p_V)\vp^{*\mu} \bar q_1\gamma_\mu(1-\gamma_5)q_2
 -\xi(\vp^*\cdot p_B)\bar q_1p\!\!\!/_V(1-\gamma_5)q_2 \non \\
&& -m_B(\vp^*\cdot p_2)\bar q_1(1+\gamma_5)q_2-\xi m_B\bar
q_1p\!\!\!/_V\vp\!\!\!/^*(1+\gamma_5)q_2 \Bigg\}, \non
\\
 A_{\rm spect}(B\to \J q_1\bar q_2) & &
=-i{G_F\over\sqrt{2}}\,{\alpha_s\over 4\pi}\,{C_F\over
N_c}\,{4\pi^2\over N_c}\,m_\J f_\J f_B(\tilde
A^t_{sp}V_{cb}V^*_{cq'}-\tilde
A^p_{sp}V_{tb}V^*_{tq'})\int^1_0d\xi\int^1_0d\drho\, \non
\\ &&\times {\Phi^\J(\xi)\Phi^B_1(\drho)\over \xi\,\drho\,(p_B\cdot
p_2)(p_\J\cdot p_2)} \Bigg\{  [p_B\cdot p_2-(\xi-r) p_B\cdot
p_\J]\vp^{*\mu} \bar q_1\gamma_\mu(1-\gamma_5)q_2 \non \\ &&
-(\xi-r)(\vp^*\cdot p_B)\bar q_1p\!\!\!/_\J(1-\gamma_5)q_2
-m_B(\vp^*\cdot p_2)\bar q_1(1+\gamma_5)q_2  \non \\ && -(\xi-r)
m_B\bar q_1p\!\!\!/_\J\vp\!\!\!/^*(1+\gamma_5)q_2 \Bigg\},
\label{Aspect}
\en
where $r=(m_cf^T_\J)/(m_\J f_\J)$. In deriving the above equation,
we have applied the on-shell conditions $\bar q_1p\!\!\!/_1=0$,
$p\!\!\!/_2q_2=0$, the approximation $\drho\approx 0$ [see the
discussion after Eq. (\ref{Bda})] and the $B$ meson wave function
\cite{BBNS1}:
\be
\la 0|\bar q_\alpha(x)b_\beta(0)|\bar
B(p)\ra\!\!\mid_{x_+=x_\bot=0}=-{if_B\over 4}[(p\!\!\!/
+m_B)\gamma_5]_{\beta\gamma}\int^1_0d\drho\, e^{-i\drho
p_+x_-}[\Phi^B_1(\drho)+n\!\!\!/_-\Phi^B_2(\drho)]_{\gamma\alpha},
 \label{Bwf}
\en
with  $n_-=(1,0,0,-1)$. The expressions for $A_{sp}^{t,p}$ and
$\tilde A_{sp}^{t,p}$ can be obtained from that of $A^{t,p}$ and
$\tilde A^{t,p}$ (see Table I) respectively by replacing the
coefficient $a_{2i}~(a_{2i-1})$ by the Wilson coefficient
$c_{2i-1}~(c_{2i})$. In passing, we note that,
in contrast to \cite{Browder}, the 2-body decay
$b\to Mq_1$ and the 3-body decay $B\to Mq_1\bar q_2$ do not
interfere with each other to give contributions to $B\to MX$.

Eq. (\ref{Aspect}) can be applied to two-body exclusive decays.
Consider $B\to K\pi$ as an example and this amounts to having
$P=K$ and a pion from $q_1\bar q_2$ . Hence $p_2=\deta p_\pi$,
where $\deta$ is the momentum fraction of the antiquark in the
pion. It follows from (\ref{Aspect}) and (\ref{tw3}) that
\be
A_{\rm spect}(B\to K\pi) \propto \int^1_0 {d\drho\over\drho}\,
\Phi^B_1(\drho)\int^1_0 {d\xi\over\xi} \,\Phi^K(\xi)\int^1_0
{d\deta\over \deta}\,\left[\Phi^\pi(\deta)+{2\mu_\chi\over
m_b}\phi^\pi_p(\deta)\right], \label{fIIt3}
\en
which was first obtained in \cite{BBNS2}. It is evident that the
terms proportional to $m_B$ in Eq. (\ref{Aspect}) give twist-3
contributions. Since the twist-3 distribution amplitude
$\Phi^\pi_p(\deta)\approx 1$, it does not vanish at the endpoints.
Consequently, there is a logarithmic divergence of the $\deta$
integral which implies that the spectator interaction in $B\to
K\pi$ decay is dominated by soft gluon exchange between the
spectator quark and quarks that form the emitted kaon, indicating
that QCD factorization breaks down at twist-3 order. The
above-mentioned infrared divergent problem does not occur in the
semi-inclusive decay, however.

The decay distribution due to hard spectator interactions is given
by
\be
d\Gamma_{\rm spect} = {1\over (2\pi)^3}\,{1\over 32m_B^3}\int
|A_{\rm spect}|^2dm_{12}^2dm_{23}^2,
\en
or
\be
{d\Gamma_{\rm spect}\over dE_M}={1\over (2\pi)^3}\,{1\over
16m_B^2}\int |A_{\rm spect}|^2dm_{23}^2,
\en
where $E_M$ is the energy of the outgoing meson $M$, and
$m_{ij}^2=(p_i+p_j)^2$ with $p_3=p_M$. For a given $E_M$, the
range of $m_{23}^2$ is fixed by kinematics. In order to
enhance the possibility that $B\to
MX$ originates from a quasi-two-body decay, we impose the energy cutoff
$E_M>2.1$ GeV for light mesons and $E_M>3.3$ GeV for the $\J$.

\subsection{Results and discussions}
To proceed we apply the initial bound state effect to hard
spectator interactions and use the $B$ meson wave function
 \be
\Phi^B_1(\drho)=N_B\drho^2(1-\drho)^2{\rm exp}\left[-{1\over
2}\left({\drho m_B\over \omega_B}\right)^2\right], \label{Bda}
 \en
with $\omega_B=0.25$ GeV and $N_B$ being a normalization constant.
This $B$ meson wave function corresponds to $\lambda_B=303$ MeV
defined by $\int^1_0 d\drho\,\Phi^B(\drho)/\drho\equiv
m_B/\lambda_B$ \cite{BBNS1}. This can be understood since the $B$
meson wave function is peaked at small $\drho$: It is of order
$m_B/\Lqcd$ at $\drho\sim \Lqcd/m_B$. Hence, the integral over
$\phi_B(\drho)/\drho$ produces an $m_B/\Lqcd$ term. As for the
parameter $r$ defined after Eq. (\ref{fIJ}), it is equal to 1/2 in
the heavy quark limit assuming $f_\J^T=f_\J$. The results of
calculations are shown in Table IV. We see that the tree-dominated
color-suppressed modes $(\pi^0,\rho^0,\omega)X_{\bar s},~\phi X$,
$\J X_s,\J X$ and the penguin-dominated mode $\omega X_{s\bar s}$
are dominated by the hard spectator corrections. In particular,
the prediction ${\cal B}(B\to \J X_s)=9.6\times 10^{-3}$ is in
agreement with the measurement of a direct inclusive $\J$
production: $(8.0\pm 0.8)\times 10^{-3}$ by CLEO \cite{JX} and
$(7.89\pm 0.10\pm 0.34)\times 10^{-3}$ by BaBar \cite{Babar}. This
is because the relevant spectator interaction is color allowed,
whereas the two-body semi-inclusive decays for these modes are
color-suppressed. As a consequence, nonfactorizable hard spectator
interactions amount to giving $a_2$ a large enhancement.

\begin{table}[htb]
\caption{$CP$-averaged branching ratios and partial-rate
asymmetries for some semi-inclusive hadronic $B$ decays with
$E_M>2.1$ GeV for light mesons and $E_\J>3.3$ GeV for the $\J$.
Branching ratios due to hard spectator interactions in the 3-body
decay $B\to Mq_1\bar q_2$ are shown in parentheses. Here $X$
denotes a final state containing no (net) strange or
charm particle, and $X_q$
the state containing the quark flavor $q$.}
\begin{center}
\begin{tabular}{ l l c  }
 Mode & BR & PRA(\%)  \\ \hline
 $\ov B^0\to \pi^-X~(\ov B^0_s\to \pi^-X_{\bar s})$ & $1.3\times 10^{-4}~(5.1\times 10^{-8})$ & -2  \\
 $\ov B^0\to \rho^-X~(\ov B^0_s\to \rho^-X_{\bar s})$ & $3.4\times 10^{-4}~(2.2\times 10^{-7})$ & -2  \\
 $\ov B^0_s\to \pi^0X_{\bar s}$ & $1.3\times 10^{-6}~(8.7\times 10^{-7})$ & 31  \\
 $\ov B^0_s\to \rho^0X_{\bar s}$ &  $4.8\times 10^{-6}~(3.7\times 10^{-6})$ & 22  \\
 $\ov B^0_s\to \omega\,X_{\bar s}$ & $5.5\times 10^{-6}~(3.4\times 10^{-6})$ & -37  \\
 $B^-\to\phi\,X$ & $2.5\times 10^{-7}~(1.9\times 10^{-7})$ & -0.5  \\
 $\ov B^0\to\pi^-X_c~(\ov B_s\to \pi^-X_{c\bar s})$ & $1.8\times 10^{-2}~(8.4\times 10^{-6})$ & 0 \\
 $\ov B^0\to\rho^- X_c~(\ov B_s\to \rho^-X_{c\bar s})$ & $4.2\times 10^{-2}~(1.1\times 10^{-4})$ & 0 \\
 $B^-\to K^0X_s$ & $3.8\times 10^{-6}~(2.9\times 10^{-9})$ & -20   \\
 $B^-\to K^{*0}X_s$ & $2.2\times 10^{-6}~(1.1\times 10^{-8})$ & -24   \\
 $\ov B^0\to K^-X~(\ov B_s\to K^- X_{\bar s})$ & $8.7\times 10^{-5}~(3.6\times 10^{-9})$ & 5   \\
 $\ov B^0\to K^{*-}X~(\ov B_s\to K^{*-} X_{\bar s})$ & $3.9\times 10^{-5}~(1.4\times 10^{-8})$ & 16  \\
 $B^-\to \ov K^0X$ & $9.7\times 10^{-5}~(7.5\times 10^{-8})$ & 0.8  \\
 $B^-\to \ov K^{*0}X$ & $5.4\times 10^{-5}~(2.9\times 10^{-7})$ & 0.9   \\
 $\ov B^0\to K^- X_c~(\ov B^0_s\to K^- X_{c\bar s})$ & $1.4\times 10^{-3}~(4.3\times 10^{-7})$ & 0 \\
 $\ov B^0\to K^{*-}X_c~(\ov B^0_s\to K^{*-} X_{c\bar s})$ & $2.3\times 10^{-3}~(4.8\times 10^{-6})$ & 0 \\
 $\ov B^0_s\to\pi^0X_{s\bar s}$ & $1.5\times 10^{-6}~(5.0\times 10^{-8})$ & 19  \\
 $\ov B^0_s\to\rho^0X_{s\bar s}$ & $4.4\times 10^{-6}~(2.2\times 10^{-7})$ & 18  \\
 $\ov B^0_s\to\omega\,X_{s\bar s}$ & $7.4\times 10^{-6}~(7.1\times 10^{-6})$ & 2  \\
 $B^-\to\phi\,X_s$ & $5.8\times 10^{-5}~(2.8\times 10^{-6})$ & 1  \\
 $B\to\J\, X_s$ & $9.6\times 10^{-3}~(9.2\times 10^{-3})$ & 0 \\
 $B\to\J\, X$ & $5.1\times 10^{-4}~(4.9\times 10^{-4})$ & 0.5  \\
\end{tabular}
\end{center}
\end{table}

It is instructive to compare the enhancement of the $\J$
production in exclusive and semi-inclusive decays. The hard
spectator diagrams denoted by $f_{II}$ have been included in the
leading-twist order calculations and it is found that $a_2(\J
K)\sim 0.06-0.05i$ \cite{CYJ,Chay}. Therefore, the real part of
$a_2(\J K)$ is enhanced by $f_{II}$, which is numerically much
larger than $f_I$, but it is still too small compared to the
experimental value $0.26\pm 0.02$ \cite{a12}. It has been shown
recently that to the twist-3 level, the coefficient $a_2(\J K)$ is
largely enhanced by the nonfactorizable spectator interactions
arising from the twist-3 kaon LCDA $\phi^K_\sigma$, which are
formally power-suppressed but chirally, logarithmically and
kinematically enhanced \cite{CYJ}. The major theoretical
uncertainty is that the infrared divergent contributions there
should be treated in a phenomenological way. In this work we found
that it is the same spectator mechanism responsible for the
enhancement observed in semi-inclusive decay $B\to \J X_s$, and
yet we do not encounter the same infrared problem as occurred in
the exclusive case, and terms proportional to $m_B$ in Eq.
(\ref{Aspect}) are not power suppressed, rendering the present
prediction more reliable and trustworthy. It is conceivable that
infrared divergences residing in exclusive decays will be washed
out when all possible exclusive modes are summed over.

It is also interesting to notice that after including the
spectator corrections, the branching ratios and PRA's for the
color-suppressed modes $\ov B^0_s\to(\pi^0,\rho^0,\omega)X_{\bar
s}$, $B^-\to\phi X$ are numerically close to that predicted in
\cite{Soni} based on naive factorization (see Table III). Note
that the large CP asymmetries in $b\to (\pi^0,\rho^0,\omega)d$
decays (see Table III) are washed out to a large extent at hadron
level by spectator interactions. By contrast, the nonfactorizable
spectator interaction is in general negligible for penguin
dominated (except for $\omega X_{s\bar s}$) or color-allowed tree
dominated decay modes. The channels $(B^-,\ov
B^0)\to(\pi^0,\rho^0,\omega,\phi)X$ are not listed in Table IV as
they involve the unwanted form factors. For example,
$B^-\to\pi^0X$ contains a term $a_2F^{B\pi}$ and $\ov
B^0\to\pi^0X$ has a contribution like $a_4F^{B\pi}$. Hence, the
prediction of $(B^-,\ov B^0)\to\pi^0X$ is not as clean as $\ov
B_s^0\to \pi^0X_{\bar s}$. Nevertheless, the former is also
dominated by spectator interactions and is expected to have the
same order of magnitude for branching ratios as the latter.

Owing to the presence of $B-\eta(\eta')$ form factors, the decays
$B\to (\eta,\eta')X$ are also not listed in Table IV.  However, we
find that the hard spectator corrections to the prompt $\eta'$
production in semi-inclusive decays are very small and hence the
four-quark operator contributions to $b\to\eta' s$ can only
account for about 10\% of the measured result, Eq. (\ref{etapX}).
Evidently this implies that one needs a new mechanism (but not
necessarily new physics) specific to the $\eta'$. It has been
advocated that the anomalous coupling of two gluons and $\eta'$ in
the transitions $b\to sg^*$ followed by $g^*\to \eta'g$ and $b\to
sg^*g^*$ followed by $g^*g^*\to\eta'$ may explain the excess of
the $\eta'$ production \cite{AS,Kagan}. An issue in this study is
about the form-factor suppression in the $\eta'-g^*-g^*$ vertex
and this has been studied recently in the perturbative QCD hard
scattering approach \cite{Ali00}. At the exclusive level, it is
well known that the decays $B^\pm\to\eta'K^\pm$ and $\ov
B^0\to\eta' \ov K^0$ have abnormally large branching ratios
\cite{PDG}. In spite of many theoretical uncertainties, it is safe
to say that the four-quark operator contribution accounts for at
most half of the experimental value and the new mechanism
responsible for the prolific $\eta'$ production in semi-inclusive
decay could also play an essential role in $B\to\eta'K$ decay.

From Table IV it is clear that the semi-inclusive decay modes:
$\ov B^0_s\to(\pi^0,\rho^0,\omega)X_{\bar s}$, $\rho^0X_{s\bar
s}$, $\ov B^0\to (K^-X,K^{*-}X)$ and $B^-\to(K^0X_s,K^{*0}X_s)$
are the most promising ones in searching for direct CP violation;
they have branching ratios of order $10^{-6}-10^{-4}$ and CP rate
asymmetries of order $(10-40)\%$. Note that as shown in Eqs.
(\ref{CPtree}) and (\ref{CPpeng}), a measurement of partial rate
difference of $\ov B^0_s\to(\pi^0,\rho^0,\omega)X_{\bar s}$ and
$B^-\to(K^0X_s,K^{*0}X_s)$ will provide useful information on the
unitarity angle $\alpha$, while $\ov B^0_s\to\rho^0X_{s\bar s}$
and $\ov B^0\to (K^-X,K^{*-}X)$ on the angle $\gamma$. To have a
rough estimate of the detectability of CP asymmetry, it is useful
to calculate the number of $B-\ov B$ pairs needed to establish a
signal for PRA to the level of three statistical standard
deviations given by
 \be
N_B^{3\sigma}={9\over \Delta^2 Br\, \epsilon_{\rm eff}},
\label{NB}
 \en
where $\Delta$ is the PRA, $Br$ is the branching ratio and
$\epsilon_{\rm eff}$ is the product of all of the efficiencies
responsible for this signal. With about $1\times 10^7$ $B\ov B$
pairs, the asymmetry in $K^{*-}$ channel starts to become
accessible; and with about $7\times 10^7$ $B\ov B$ events, the
PRA's in the other modes mentioned above will become feasible.
Here we assumed, for definiteness, $\epsilon_{\rm eff}=1$ and a
statistical significance of 3$\sigma$ as in Eq. (\ref{NB}).
Currently BaBar has collected 23 million $B\ov B$ events, BELLE 11
million pairs and CLEO 9.6 million pairs. It is conceivable that
CP asymmetries in semi-inclusive $B$ decays will begin to be
accessible at these facilities. Likewise, PRA's in semi-inclusive
$B_s$ decays may be measurable in the near future at the
Fermilab's Tevatron.

It is interesting to note that the decays $\ov B^0_s\to
(\pi^0,\rho^0,\omega)X_{s\bar s}$ and $B^-\to\phi X$ are
electroweak-penguin dominated. Except for the last channel, they
have sizable branching ratios and two of them have observable CP
asymmetries. A measurement of these reactions will provide a good
probe of electroweak penguins.

Finally, it is useful to discuss briefly the theoretical
uncertainties one may have in the present approach for
semi-inclusive decays: the $b$ quark mass, the annihilation
diagram and the distribution amplitude of the meson. We have
assumed a pole mass for the $b$ quark to compute the decay rates
of $b\to Mq$ and $B\to MX$ which are proportional to $m_b^3$. In
principle, this uncertainty in $m_b$ can be reduced by normalizing
the semi-inclusive hadronic rate to the semi-leptonic one. Since
the latter is of 5th power in $m_b$, the uncertainty is only
slightly alleviated. The annihilation topology is power suppressed
in the heavy quark limit and is conventionally assumed to be
small. However, it is conceivable that power corrections due to
the annihilation diagrams could be important for penguin-dominated
semi-inclusive decays such as $B^-\to \ov K^0(\ov K^{*0})X$ and
$\ov B^0\to K^-(K^{*-})X$. As for the LCDAs of the meson, we have
assumed the asymptotic form (\ref{LCDA}) for the leading-twist
LCDA, which is suitable for light mesons but probably not so for
the heavy meson $\J$. Due to SU(3) symmetry breaking, the
realistic kaon wave function should exhibit a slight asymmetry in
$1-2x$. Also the distribution amplitude of the $B$ meson is not
well understood; phenomenologically, the parameter $\omega_B$ [see
Eq. (\ref{Bda})] or $\lambda_B$ is not well fixed.

\section{Conclusions}
We have systematically investigated semi-inclusive $B$ decays
$B\to MX$ within the framework of QCD-improved factorization. The
nonfactorizable effects, such as vertex-type and penguin-type
corrections to the two-body $b$ decay, $b\to Mq$, and hard
spectator corrections to the 3-body decay $B\to Mq_1\bar q_2$ are
calculable in the heavy quark limit. QCD factorization is
applicable when the emitted meson is a light meson or a
charmonium.

There are two strong phases in the QCD factorization approach: one
form final-state rescattering due to hard gluon exchange between
$M$ and $X$, and the other from the penguin diagrams. We have
discussed the issue of  the CPT constraint on partial rate
asymmetries. The strong phase coming from final-state rescattering
due to hard gluon exchange between the final states $M$ and $X$
[see Eq. (\ref{fI})] can induce large rate asymmetries for
tree-dominated color-suppressed modes
$(\pi^0,\rho^0,\omega)X_{\bar s}$. The predicted coefficient $a_2$
in QCD factorization is very small compared to naive
factorization. Consequently, the color-suppressed modes
$(\pi^0,\rho^0,\omega)X_{\bar s},~\phi X$ and $\J X_s,\J X$ are
very suppressed. Fortunately, the nonfactorizable hard spectator
interactions in $B\to Mq_1\bar q_2$, though phase-space
suppressed, are extremely important for the aforementioned modes.
In fact, they are dominated by the hard spectator corrections.
This is because the relevant hard spectator correction is color
allowed, whereas the two-body semi-inclusive decays for these
modes are color-suppressed. Our prediction ${\cal B}(B\to \J
X_s)=9.6\times 10^{-3}$ is in agreement with experiment. Contrary
to the exclusive hadronic decay, the spectator quark corrections
here are not subject to the infrared divergent problem, rendering
the present prediction more clean and reliable.

Owing to the destructive QCD anomaly effect in the matrix element
of pseudoscalar densities for $\eta'$-vacuum transition, the
four-quark operator contribution to $b\to\eta's$ is too small to
explain the observed fast $\eta'$ production. It is evident that a
new mechanism such as the anomalous coupling of two gluons with
the $\eta'$ is needed in order to resolve the $\eta'$ puzzle.

The semi-inclusive decay modes: $\ov
B^0_s\to(\pi^0,\rho^0,\omega)X_{\bar s}$, $\rho^0X_{s\bar s}$,
$\ov B^0\to (K^-X,K^{*-}X)$ and $B^-\to(K^0X_s,K^{*0}X_s)$ are the
most promising ones in searching for direct CP violation; they
have branching ratios of order $10^{-6}-10^{-4}$ and CP rate
asymmetries of order $(10-40)\%$. With about $7\times 10^7$ $B\ov
B$ pairs, CP asymmetries in these modes may be measurable in the
near future at the BaBar, BELLE, CLEO and Tevatron experiments.
The decays $\ov B_s^0\to (\pi^0,\rho^0,\omega)X_s$ and $B^-\to\phi
X$ are electroweak-penguin dominated. Except for the last mode,
they in general have sizable branching ratios and two of them have
observable CP asymmetries. The above-mentioned reactions will
provide good testing ground for the standard model and a good
probe for electroweak penguins.

\vskip 2.0cm \acknowledgments We thank David Atwood for discussions.
One of us (H.Y.C.) also wishes to thank
Physics Department, Brookhaven National Laboratory for its
hospitality. This work was supported in part by the U.S. DOE
contract No. DE-AC02-98CH10886(BNL) and by the National Science
Council of R.O.C. under Grant No. NSC89-2112-M-001-082.


\end{document}